\documentclass[
  ,draft            
  ]
  {aipproc}

\layoutstyle{6x9}

\begin{document}

\title{Electromagnetic Form Factors of Hadrons in Quantum Field Theories}

\author{C. A. Dominguez}{
  address={Centre for Theoretical Physics and Astrophysics\\
University of Cape Town, Rondebosch 7700, South Africa, and Department of Physics, Stellenbosch University, Stellenbosch 7600, South Africa}
}

\begin{abstract}
In this talk, recent results are presented of calculations of electromagnetic form factors of hadrons  in the framework of two quantum field theories (QFT), (a) Dual-Large $N_c$ QCD (Dual-$QCD_\infty$) for the pion, proton, and $\Delta(1236)$, and (b) the Kroll-Lee-Zumino (KLZ) fully renormalizable Abelian QFT for the pion form factor. Both theories provide a QFT platform to improve on naive (tree-level) Vector Meson Dominance (VMD). Dual-$QCD_\infty$ provides a tree-level improvement by incorporating an infinite number of zero-width resonances, which can  be subsequently shifted  from the real axis to account for the time-like behaviour of the form factors. The renormalizable KLZ model provides a QFT improvement of VMD in the framework of perturbation theory. Due to the relative mildness of the $\rho\pi\pi$ coupling, and the size of loop suppression factors, the perturbative expansion is well defined in spite of this being a strong coupling theory. Both approaches lead to considerable improvements of VMD predictions for electromagnetic form factors, in excellent agreement with data. 
\end{abstract}

\maketitle

\section{Introduction}
In its original formulation \cite{SAKU}, Vector Meson Dominance (VMD) is an effective tree-level model based on the notion of $\gamma-\rho^0$ conversion. When applied to e.g. electromagnetic form factors, it can roughly account for the pion form factor in the space-like region, and with some modifications,  also in the time-like region around the rho-meson peak. However, for non-zero spin hadrons such as nucleons and $\Delta(1236)$, VMD is in serious disagreement with the observed $q^2$ fall-off of these form factors. This situation can hardly be remedied without a  dynamical platform allowing to go beyond naive, single pole, tree-level in a systematic fashion, i.e. a renormalizable QFT framework. An attempt in this direction was made long ago by incorporating radial excitations of the rho-meson into VMD, i.e. Extended VMD \cite{EVMD}. At the time, however, there was no known renormalizable QFT to support this approach. Today, we know that in the limit of an infinite number of colours, QCD is solvable  leading to a hadronic spectrum consisting of an infinite number of zero-width states \cite{QCDINF}. Unfortunately, the masses and couplings of these states remain unspecified, so that models are needed to fix these parameters. An attractive and highly economical candidate (in terms of free parameters) is Dual-$QCD_\infty$ \cite{CAD1}-\cite{CAD3}, {\bf inspired} in the Dual Resonance Model for scattering amplitudes of Veneziano \cite{VEN}, the precursor of string theory. It is very important to stress the word {\bf inspired}, as Dual-$QCD_\infty$ does not share any of the unwanted features of the original Veneziano model, such as lack of unitarity, unphysical particles in the spectrum, etc. In fact, in Dual-$QCD_\infty$  the masses and couplings of the zero-width states are fixed so that form factors become Euler Beta functions, involving one single free parameter which controls their asymptotic power behaviour. This remains the only connection between Dual-$QCD_\infty$, which so far has only been applied to three-point functions,  and the Dual Resonance Model originally formulated for n-point functions ($n \geq 4$). Another aspect of Dual-$QCD_\infty$ which needs to be stressed, to avoid misunderstandings, is that it is not intended to be an expansion in powers of $1/N_c$ \cite{CAT}. In fact, $N_c$ is taken to be infinite from the start, as this is the limit in which QCD is solvable and leads to the hadronic spectrum mentioned above. Unitarization can subsequently be performed by shifting the poles from the real axis into the second Riemann sheet in the complex energy (squared) plane. This induces corrections to form factors of order $\cal{O}$$(\Gamma/M \simeq 10\%)$. But essentially Dual-$QCD_\infty$ remains a tree-level QFT improvement over VMD.\\

Another highly attractive improvement of tree-level VMD can be achieved in the framework of the Kroll-Lee-Zumino QFT of pions and a massive (neutral) rho-meson \cite{KLZ}. In spite of the presence in the KLZ Lagrangian of an explicit mass term for the rho-meson, this theory is perfectly renormalizable as long as the gauge field remains Abelian \cite{KLZ}. The great advantage of a renormalizable QFT is the absence of free parameters. However, since in this case we are dealing with a strong coupling theory, it is essential to have a meaningful perturbative expansion. This has been shown to be the case for the pion form factor in the time-like \cite{GK}, as well as the space-like region \cite{CAD4}. This is due to the relative smallness of the $\rho\pi\pi$ coupling, and the large loop suppression factors. An extension of this theory to include vector meson radial excitations is certainly possible, and would establish an interesting connection with Dual-$QCD_\infty$. It would also extend the momentum transfer (time-like) region of validity of the calculated pion form factor.\\ 

\section{Dual-$QCD_\infty$}
In $QCD_\infty$, a typical form factor has the generic form
\begin{equation}
F(s) = \sum_{n=0}^{\infty}
\frac{C_{n}}{(M_{n}^{2} -s)} \;,
\end{equation}
where $s \equiv q^2$ is the momentum transfer squared, and  the masses $M_n$, and the couplings $C_n$ remain unspecified. In Dual-$QCD_{\infty}$ they are given by
\begin{equation}
C_{n} = \frac{\Gamma(\beta-1/2)}{\alpha' \sqrt{\pi}} \; \frac{(-1)^n}
{\Gamma(n+1)} \;
\frac{1}{\Gamma(\beta-1-n)} \;, 
\end{equation}
where $\beta$ is a free parameter, and the string tension $\alpha'$ is $\alpha' = 1/2 M_{\rho}{^2}$,  as it enters the rho-meson Regge trajectory
$\alpha_{\rho}(s) = 1 + \alpha ' (s-M_{\rho}^{2})$. The mass spectrum is chosen as
$M_{n}^{2} = M_{\rho}^{2} (1 + 2 n)$. This simple formula correctly predicts the first few radial excitations. Other, e.g. non-linear mass formulas could be used, but this hardly changes the results in the space-like region, and only affects the time-like region behaviour for very large $q^2$. With these choices the form factor becomes an Euler Beta-function, i.e.
\begin{eqnarray}
F(s) &=& \frac{\Gamma(\beta-1/2)}{\sqrt{\pi}} \; \sum_{n=0}^{\infty}\;
\frac{(-1)^{n}}{\Gamma(n+1)} \; \frac{1}{\Gamma(\beta-1-n)}\frac{1}
{[n+1-\alpha_\rho(s)]} \nonumber\\ [.5cm]
& = &
\frac{1}{\sqrt{\pi}} \; \frac{\Gamma (\beta-1/2)}{\Gamma(\beta-1)}   \;\;
B(\beta - 1,\; 1/2 - \alpha' s)\;,
\end{eqnarray}
where $B(x,y) = \Gamma(x) \Gamma(y)/\Gamma(x+y)$. The form factor exhibits asymptotic power behaviour in the space-like region, i.e.
\begin{equation}
\lim_{s \rightarrow - \infty} F(s) = (- \alpha' \;s)^{(1-\beta)}\;,
\end{equation}
from which one identifies the free parameter $\beta$ as controlling this asymptotic behaviour. Notice that while each term in Eq.(3) is of the monopole form, the result is not necessarily of this form because it involves a sum over an infinite number of states. The exception occurs for integer values of $\beta$, which leads to a finite sum.
\begin{figure}
  \includegraphics[height=.4\textheight]{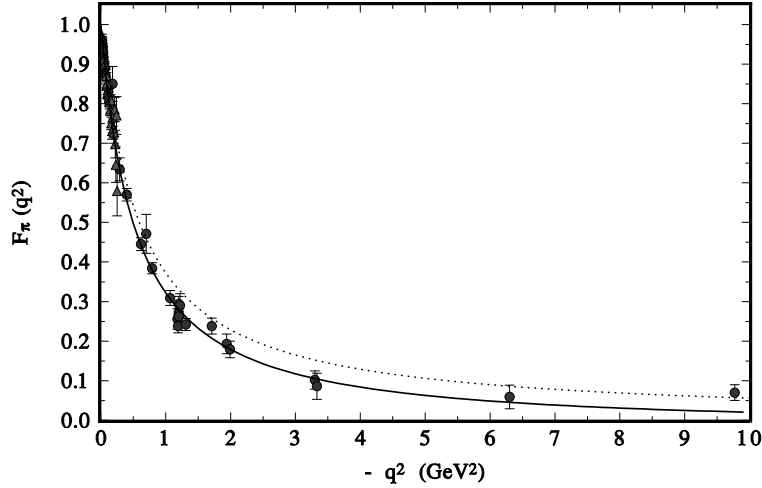}
  \caption{Pion form factor in Dual-$QCD_\infty$ (solid curve), which overlaps with the KLZ result,  and VMD (broken curve).}
  \label{fig:figure1}
\end{figure}
The imaginary part of the form factor Eq.(3) is
\begin{equation}
Im \; F(s) = \frac{\Gamma(\beta-1/2)}{\alpha' \sqrt{\pi}} \;
\sum_{n=0}^{\infty} \; \frac{(-1)^{n}}{\Gamma(n+1)} 
 \frac{1}{\Gamma(\beta-1-n)} \; \pi \; \delta(M_n^2-s) \;.
\end{equation}
Unitarization can be performed by shifting the poles from the real axis in the complex s-plane. The simplest model is the Breit-Wigner form
\begin{equation}
\pi \delta(M_{n}^{2} -s) \rightarrow \frac{\Gamma_{n} M_{n}}
{[(M_{n}^{2} - s)^{2} + \Gamma_{n}^{2} M_{n}^{2}]} \;,
\end{equation}
where one expects $\Gamma_n$ to grow with $M_n$. Other, more refined choices, are certainly possible, e.g. the Gounaris-Sakurai  form in which the width is momentum transfer dependent.\\
\begin{figure}[ht]
\begin{minipage}[b]{0.5\linewidth}
\includegraphics[scale=0.5]{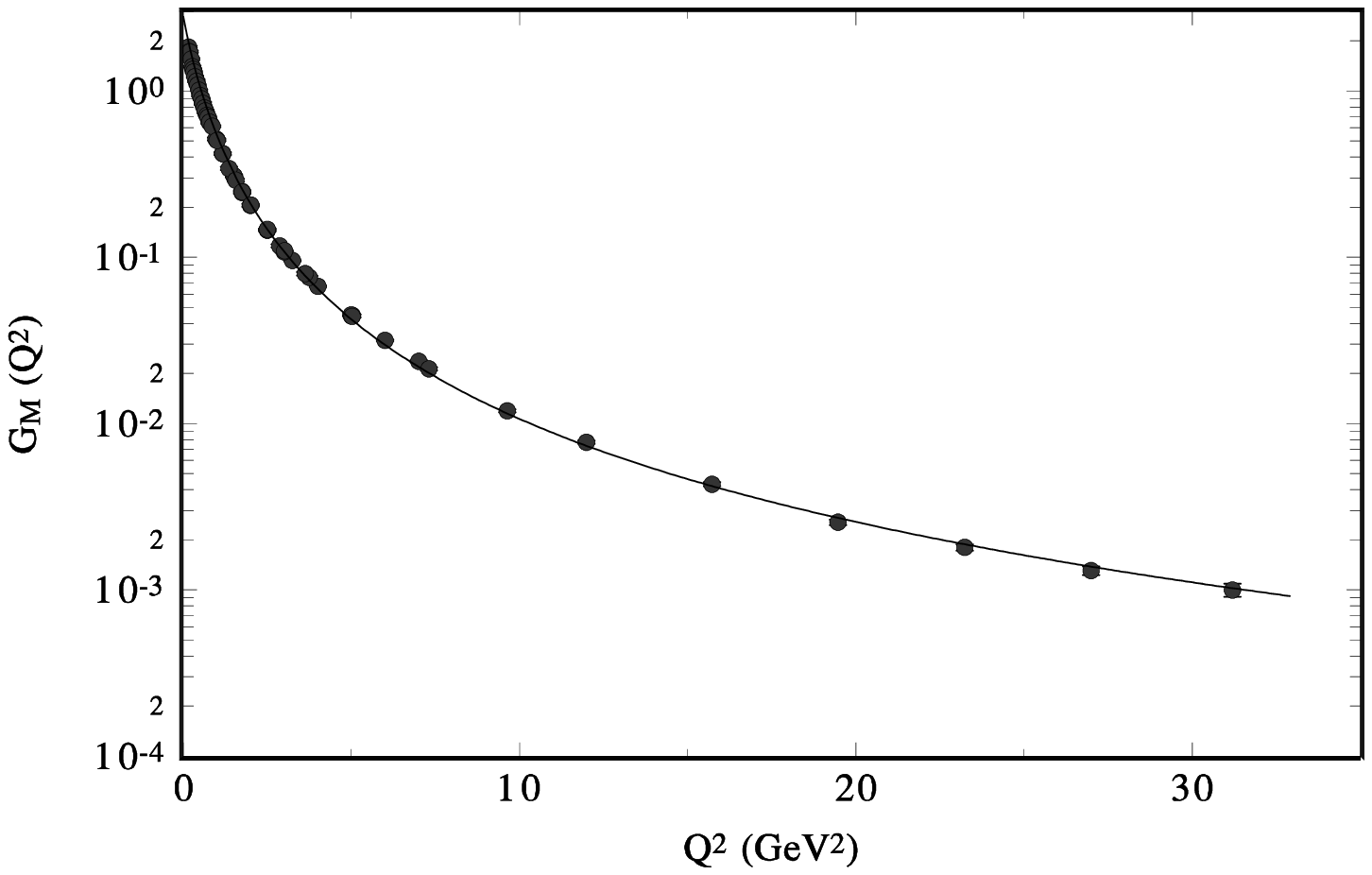}
\caption{default}
\label{fig:figure2}
\end{minipage}
\hspace{0.5cm}
\begin{minipage}[b]{0.5\linewidth}
\includegraphics[scale=0.60]{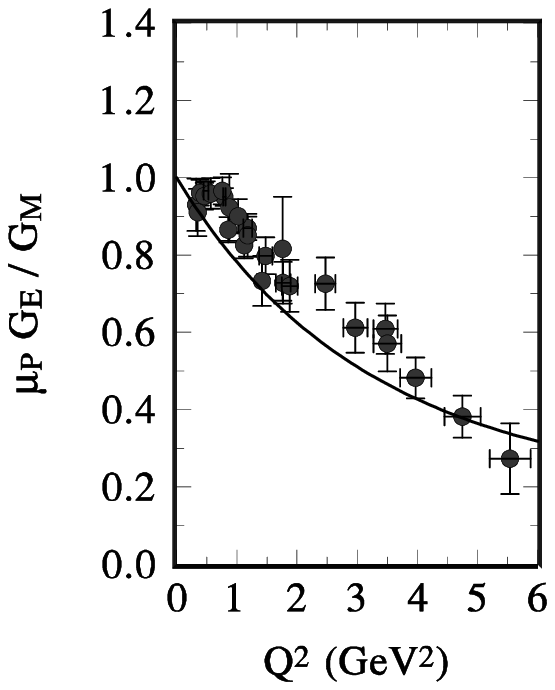}
\caption{Sachs magnetic form factor $G_M(Q^2)$ (left), and ratio of electric to magnetic Sachs form factors (right). Data is from the reanalysis of \cite{NDATA}, and $Q^2\equiv - q^2$.}
\label{fig:figure2}
\end{minipage}
\end{figure}
Space-like results in this framework are shown, together with the data, in Fig. 1 for the pion (data from \cite{DATAPI}), Fig. 2 for the proton, and Figs. 3 and 4  for the $\Delta(1236)$. In the case of the proton, Eq.(3) is used for the Dirac and Pauli form factors, $F_1(q^2)$ and $F_2(q^2)$, as these have the correct analyticity properties. Once fitted to the data, the Sachs form factors $G_E(q^2)$ and $G_M(q^2)$ follow. The fits have been made not to the raw data, but rather to the  data base as corrected in \cite{NDATA}. These corrections take into account the discrepancies between unpolarized (SLAC) and polarized (JLAB) experiments. For the $\Delta(1236)$, the three so called Scadron form factors $G^*_{M,E,C}(q^2)$ were fitted using Eq.(3), and data on $G^*_M(q^2)$, and the two ratios between $G^*_{E,C}(q^2)$ and $G^*_M(q^2)$ \cite{DATA1}-\cite{DATA2}. The value of the free parameter $\beta$ in the form factor, Eq. (3), which determines its asymptotic behaviour, is as follows: for the pion, $\beta_\pi = 2.3$, for $F_1$ and $F_2$ of the proton, $\beta_1 = 2.95 - 3.03$, and $\beta_2=4.13 - 4.20$, and for $G^*_M$, $G^*_E$, $G^*_C$ of the $\Delta(1236)$, $\beta^*_M = 4.6 - 4.8$, $\beta^*_E \simeq \beta^*_M$, and $\beta^*_C = 6.0 - 6.2$. Taking the middle values of these numbers, the asymptotic behaviour in the space-like region of these form factors is approximately as follows: $F_\pi \sim (- q^2)^{- 1.3}$,
$F_1 \sim (- q^2)^{- 2.0}$, $F_2 \sim (- q^2)^{- 3.2}$, $G^*_M \sim G^*_E \sim (- q^2)^{- 3.7}$, and $G^*_C \sim (- q^2)^{- 5.1}$. The pion form factor has also been determined in the time-like region using the simple unitarization procedure as in Eq. (6). In spite of the simplicity of the model, the result is in good agreement with data at and around the $\rho$ peak \cite{CAD1}. Proton form factors in the time-like region are currently under study, together with neutron form factors \cite{CAD5}.
\begin{figure}[ht]
  \includegraphics[height=.35\textheight]{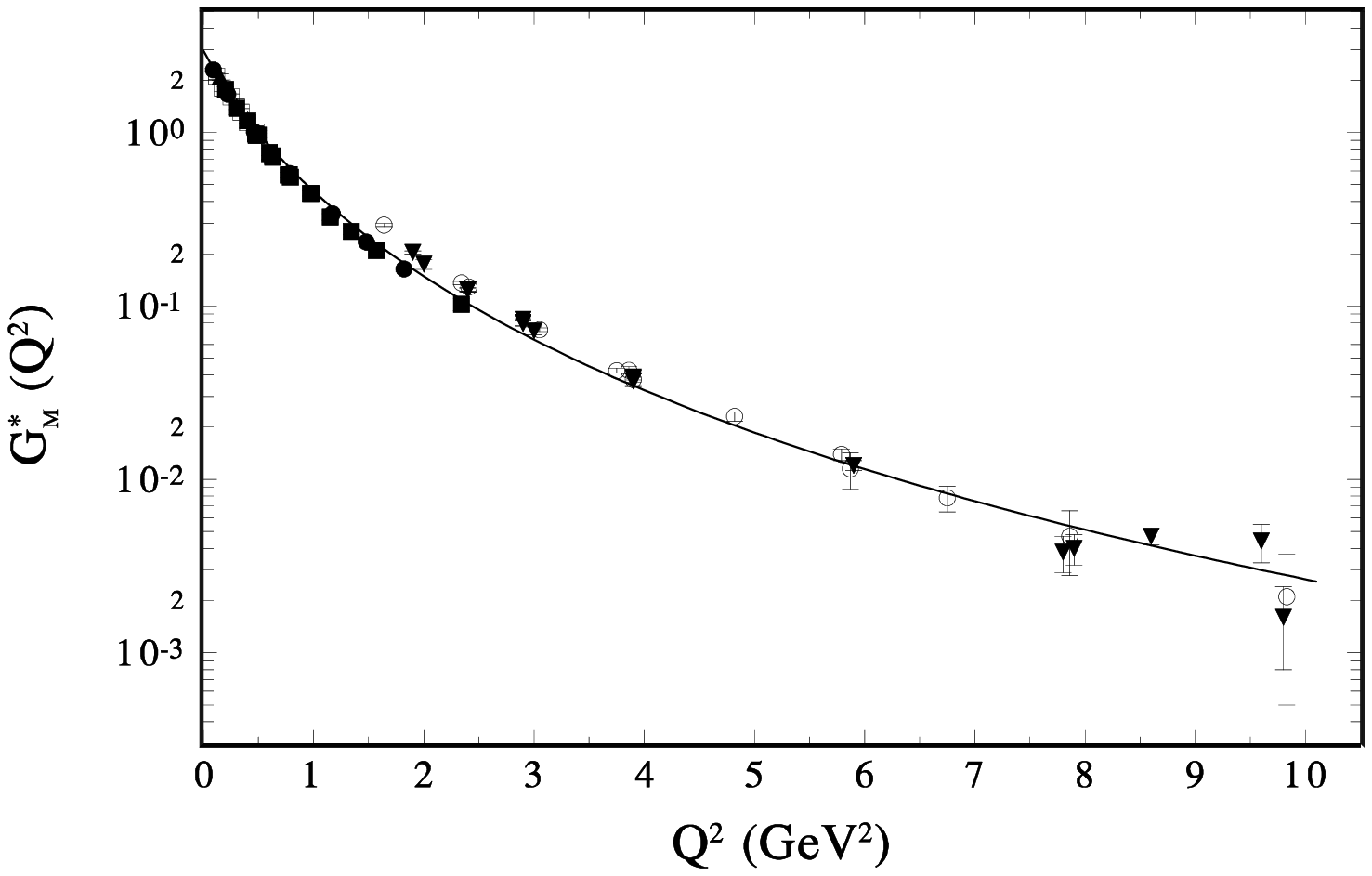}
  \caption{$\Delta(1236)$ magnetic form factor in Dual-$QCD_\infty$ (solid line).}
\label{fig:figure3}
\end{figure}
\begin{figure}[ht]
  \includegraphics[height=.35\textheight]{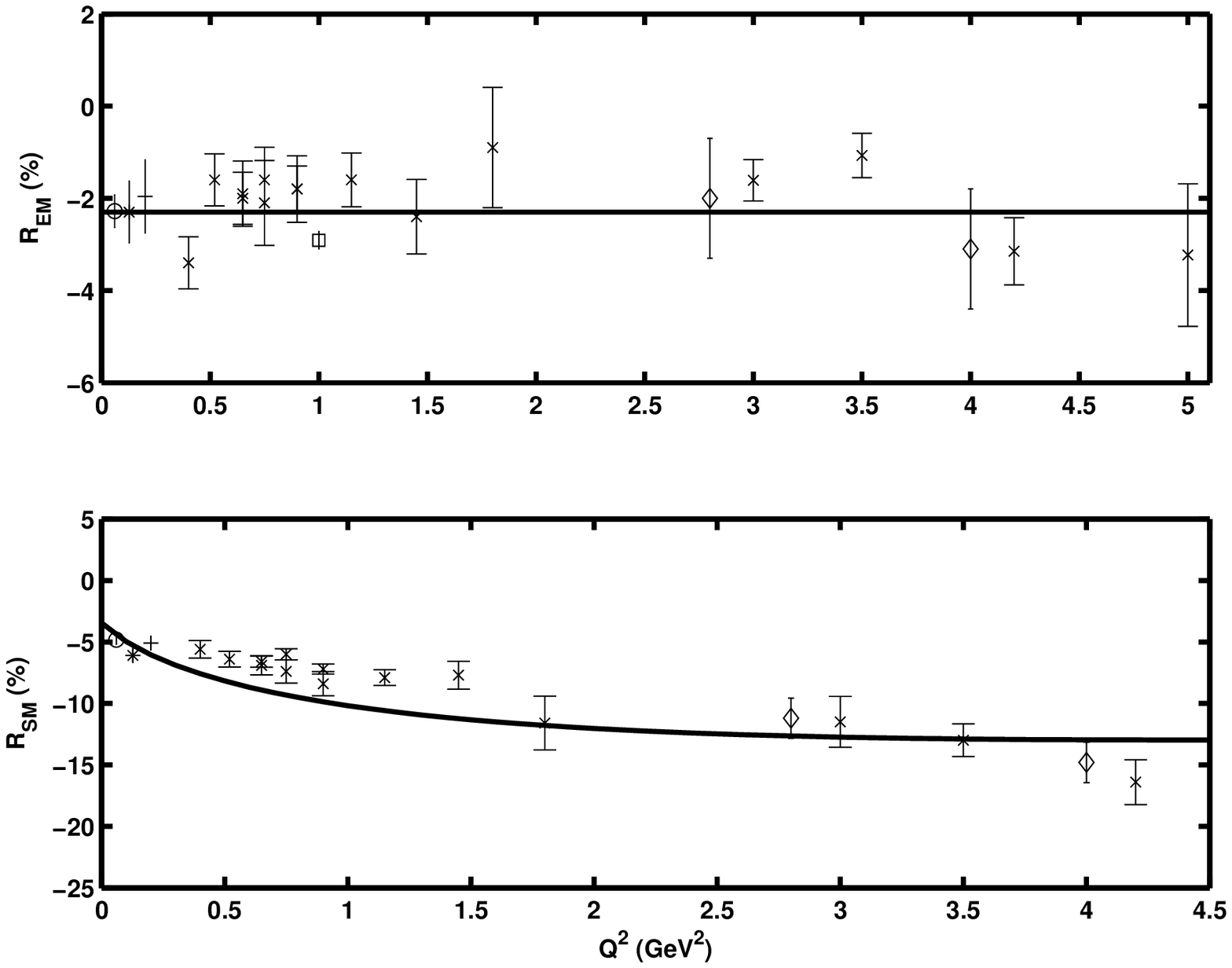}
  \caption{$\Delta(1236)$ ratios of electric to magnetic and Coulomb to magnetic form factors in Dual-$QCD_\infty$ (solid line).}
\label{fig:figure4}
\end{figure}

\section{Kroll-Lee-Zumino Quantum Field Theory}
The KLZ theory is defined by the Lagrangian \cite{KLZ}
\begin{equation}
\mathcal{L}_{KLZ} = \partial_\mu \phi \; \partial^\mu \phi^* -  m_\pi^2 \;\phi \;\phi^* - \frac{1}{4}\; \rho_{\mu\nu} \;\rho^{\mu\nu} + \frac{1}{2}\; m_\rho^2\; \rho_\mu \;\rho^\mu 
+g_{\rho\pi\pi} \rho_\mu J^\mu_\pi\;,
\end{equation}
where $\rho_\mu$ is a vector field describing the $\rho^0$ meson ($\partial_\mu \rho^\mu = 0$), $\phi$ is a complex pseudo-scalar field describing the $\pi^\pm$ mesons, $\rho_{\mu\nu}$ is the usual field strength tensor, and $J^\mu_\pi$ is the $\pi^\pm$ current.
Omitted from Eq.(7)  is an additional term of higher order in the coupling, of the form $g_{\rho\pi\pi}^2 \;\rho_\mu\; \rho^\mu \;\phi \;\phi^*$, which is not relevant to the present work. In spite of the explicit mass term for the rho-meson in this Lagrangian, this QFT has been shown to be renormalizable. This is due to the fact that the neutral vector mesons are coupled only to conserved currents.
At leading order in perturbation theory (tree-level) KLZ reproduces VMD predictions. However, at next-to-leading order and beyond, it provides a  QFT framework to systematically calculate  corrections to VMD. Because of renormalizability, these corrections do not involve free parameters (the masses and couplings in the Lagrangian are known from experiment).
\begin{figure}[ht]
  \includegraphics[height=.1\textheight]{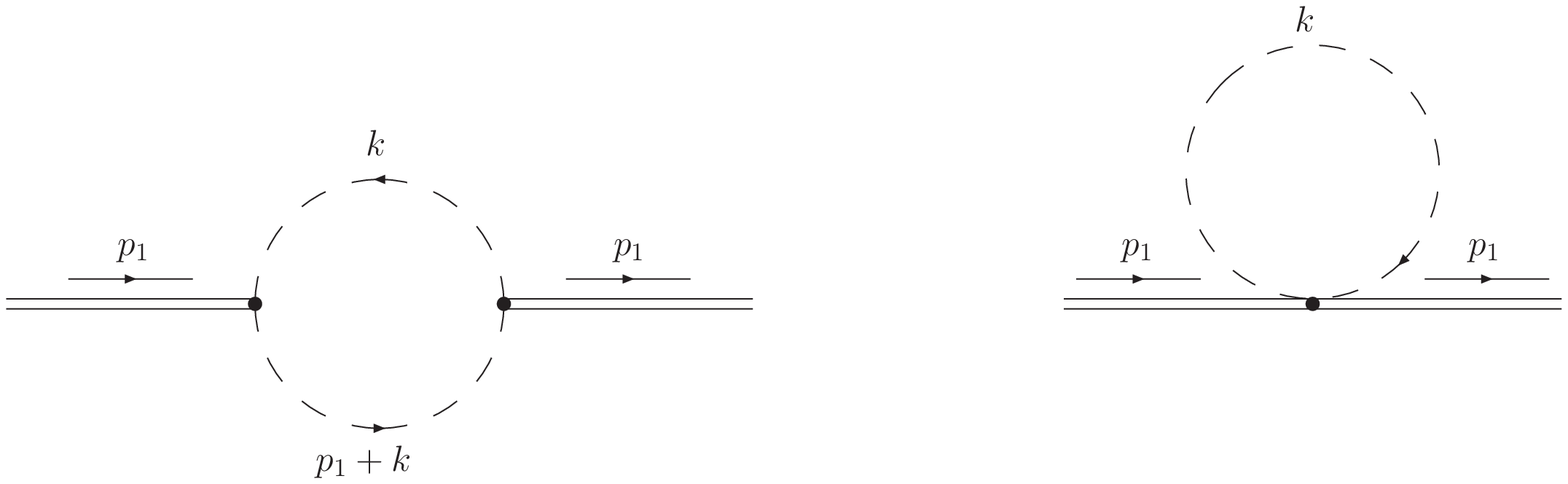}
  \caption{Vacuum polarization in KLZ.}
\label{fig:figure5}
\end{figure}
\begin{figure}[ht]
  \includegraphics[height=.15\textheight]{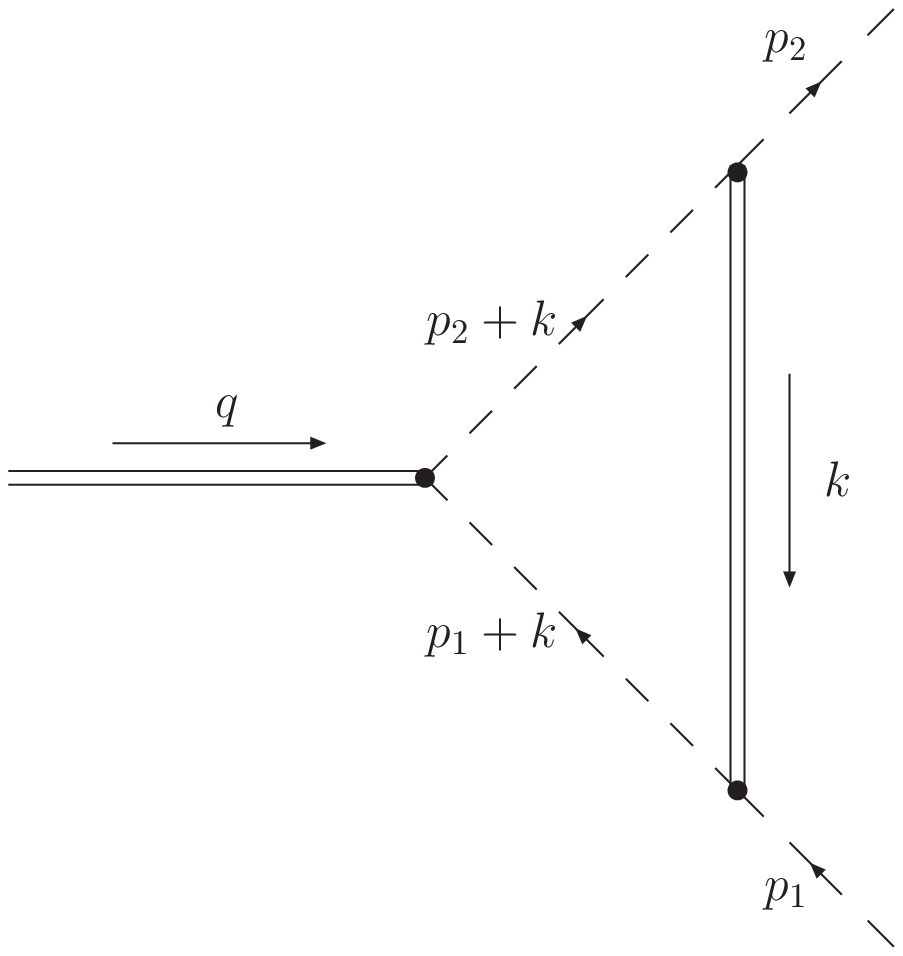}
  \caption{Vertex diagram in KLZ.}
\label{fig:figure6}
\end{figure}
The one-loop level corrections due to vacuum polarization, Fig.5, have been calculated in \cite{GK}, and those due to the vertex diagram, Fig.6, in \cite {CAD4}. These calculations were performed in dimensional regularization, and subsequently a standard renormalization procedure was followed. Vacuum polarization has been renormalized on mass shell ($q^2 = M_\rho^2$), and vertex renormalization at $q^2=0$ in order to make use of the known value of the pion form factor $F_\pi(0) = 1$. This form factor is given by
\begin{equation}
F_\pi(q^2) = \frac{M_\rho^2 + \Pi(0)|_{\mbox{vac}}}{M_\rho^2 - q^2 + \Pi(q^2)|_{\mbox{vac}}} + \frac{M_\rho^2}{M_\rho^2 - q^2} \Big[ G(q^2) - G(0)\Big] \;,
\end{equation} 
where $\Pi(q^2)|_{\mbox{vac}}$ stands for the vacuum polarization contribution (see \cite{GK}), and $G(q^2)$ is the vertex correction, both terms being of order $\cal{O}$($g^2_{\rho\pi\pi}$) (for details see \cite{CAD4}). 
\begin{figure}[ht]
  \includegraphics[height=.3\textheight]{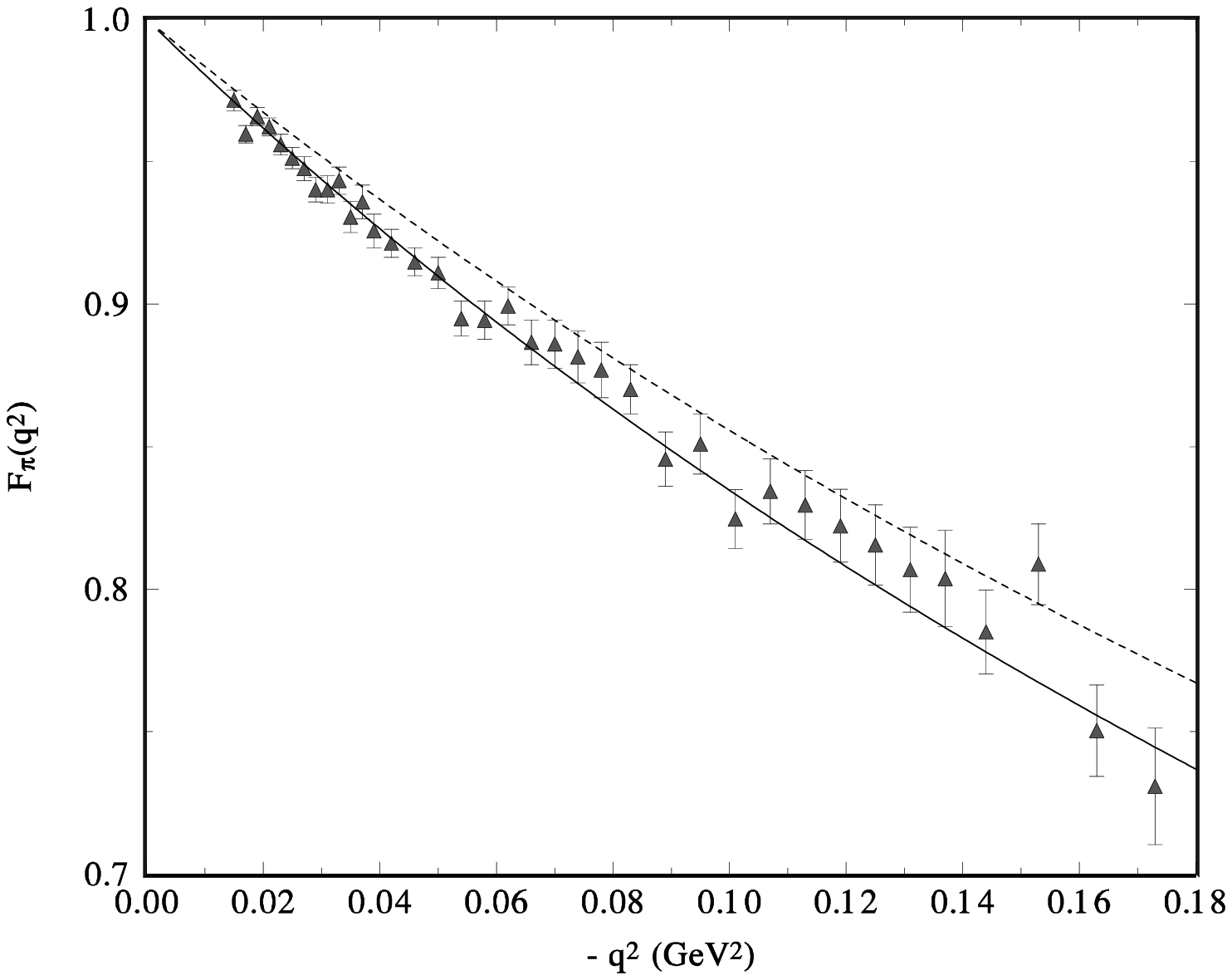}
  \caption{Pion form factor at low $q^2$ from KLZ (solid line), and from VMD (dotted line). The KLZ form factor basically overlaps with that from Dual-$QCD_\infty$, Fig.1. }
\label{fig:figure7}
\end{figure}
Results for $F_\pi(q^2)$ in the space-like region essentially overlap with those of Dual-$QCD_\infty$, Fig.1. To highlight the differences with VMD, Fig.7 shows the low $q^2$ data together with the KLZ and VMD form factors. Turning to the time-like region, at the one-loop level, i.e. to order $\cal{O}$$(g^2_{\rho\pi\pi})$, the vacuum polarization correction does not appear in the second term of Eq.(8). At and near the rho-meson peak, the width $\Gamma_\rho$ (identifiable from the imaginary part of $\Pi(q^2)$) exhibits a momentum transfer dependence, i.e.
\begin{equation}
\Gamma_\rho(s)|_{KLZ} = \frac{M_\rho\, \Gamma_\rho}{\sqrt{s}} \Big[\frac{s - 4\, \mu_\pi^2}{M_\rho^2 - 4\, \mu_\pi^2}\Big]^{\frac{3}{2}} \;,
\end{equation}
with $s \equiv q^2$. This is precisely the momentum dependent Gounaris-Sakurai (GS) width, 
known to provide an excellent fit to the data in this region. This is a rather intriguing feature, as it follows automatically from $\Pi(q^2 = M^2_\rho)|_{\mbox{vac}}$, while the GS width is a purely empirical fit formula. A very important feature of KLZ is that the one-loop corrections to tree-level VMD turn out to be  small, in spite of KLZ being a strong coupling theory. This is due to the relative mildness of the $\rho\pi\pi$ coupling ($g_{\rho\pi\pi} \sim 5$), and the large loop suppression factor $1/(4 \pi)^2$. This fortunate circumstance guarantees a meaningful perturbative expansion.

\section{Conclusions}
In this talk I have reviewed two QFT frameworks in which to compute corrections to VMD results for electromagnetic form factors of hadrons. The first is Dual-$QCD_\infty$, a Dual Resonance Model inspired realization of QCD in the limit of an infinite number of colours. This is a tree-level improvement of naive (single rho-meson) VMD, which incorporates an infinite number of vector meson radial excitations. Due to this infinite number of states, the form factor is no longer restricted to an asymptotic monopole type of behaviour. In fact, since the form factor becomes an Euler Beta function, its asymptotic behaviour is given by Eq.(4). This feature is essential to account for the fact that the pion form factor deviates slightly from a monopole, while the nucleon and $\Delta(1236)$ form factors show a very strong deviation. Dual-$QCD_\infty$ form factors involve a single free parameter in the space-like region, and at least one more in the time-like region (after unitarization). The second platform is the KLZ renormalizable Abelian gauge QFT of pions and a neutral rho-meson. This allows for a systematic calculation of corrections to tree-level VMD in the framework of perturbation theory. The perturbative expansion is meaningful, in spite of the strong coupling nature of the theory, due to the relative mildness of the $\rho\pi\pi$ coupling and to large loop suppression factors. An added advantage is that KLZ involves no free parameters on account of renormalizability.\\
Results from these two frameworks are in excellent agreement with experimental data for the pion, proton, and $\Delta(1236)$. 

\begin{theacknowledgments}
This talk is based on work done by the author, and in collaboration with J.~I. Jottar, M. Loewe, R. R\"{o}ntsch, T. Thapedi, and B. Willers. Work supported in part by the NRF (South Africa). The author thanks the organizers for  an interesting and fruitful conference. 
\end{theacknowledgments}

\end{document}